# Numerical study of the degree of light scattering strength versus fractal dimension in optical fractal disordered media: Applications in strong to weak disordered media

ETHAN R. AVERY, PEEYUSH SAHAY, SHIRSENDU NANDA, BINOD REGMI, AND PRABHAKAR PRADHAN*

*Department of Physics and Astronomy, Mississippi State University, Mississippi State, MS 39762, USA*
*Corresponding author: pp838@msstate.edu*

**Optical scattering strength of fractal optical disordered media with varying fractal dimension is reported. The diffusion limited aggregation (DLA) technique is used to generate fractal samples in 2D and 3D, and fractal dimensions are calculated using the box-counting method. The degree of structural disorder of these samples are calculated using their light localization strength, using the inverse participation ratio (IPR) analyses of the optical eigenfunctions. Results show non-monotonous behavior of the disorder-induced scattering strength with the fractal dimension, attributed to the competition between increasing structural disorder due to decrease in fractality versus decrease in scattering centers due to decreasing fractality.**

http://dx.doi.org/10.1364/OL.99.099999

Anderson postulated disorder-induced wave localization due to multiple scattering and interference effects in disordered media [1]. Since then, wave localization in disordered systems has been a subject of intense investigation, including the light wave in disordered optical media. It is also known that disordered systems with a self-similar structure can be characterized by fractal geometry. Naturally occurring porous materials, biological systems, etc., are considered as fractal in nature. A fractal system is defined by its spatial mass density correlation decay, follows as a power law in space, whose non-integer exponent is termed as the 'fractal dimension ($D_f$)' of the system [2]. Fractal geometry analysis, in terms of fractal dimension, is often employed to characterize a wide range of naturally occurring optical disordered systems, such as aerosol clusters [3], biological samples [4–6], etc. The optical scattering from a refractive index (*RI*) fluctuating ($n(r)$) medium can be characterized by onsite fluctuations of $n(r)$ and its spatial correlation length $l_c$, therefore, we expect scattering properties of fractal disordered media to depend strongly on fractality, i.e., spatial structural geometry/correlation of these fractal systems. However,
the relationship between the fractal dimension of a system and its light scattering properties is not well studied. When a solid system (with a starting integer dimension: 1 or 2 or 3) changes to a fractal system due to the development of the appropriate nature of porosity, the disorder properties of the system start increasing at the beginning. However, with the increase of porosity, the scattering properties of the system goes through a competition: (i) increasing structural disorder due to the decrease in fractality (or increased porosity) versus (ii) decrease (loss) in the scattering centers due to the decreased fractality (or increased porosity). Due to this opposite competition with the increase of porosity, the degree of scattering strength due to the disorder may not be a monotonous function with the change in fractal dimension, and one expects optimal scattering point(s). This problem is important to many branches of optical condensed matter systems, where changes in fractal dimensions are used for sample characterizations. For instance, measuring the change in the fractal dimension can be used to detect the progress of cancer in tissues [7,4,8,9]. Therefore, the exact nature of the scattering with the fractal dimension variation demands a comprehensive study to locate the optimal scattering point(s) and to characterize how the scattering properties of the system change with the change in fractal dimension.

In this work, we numerically investigate the relationship between the degree of structural disorder and fractal dimension based on their degree of light localization properties, for 2D and 3D fractal disordered optical systems. 2D and 3D fractal optical systems are generated by diffusion limited aggregation (*DLA*) simulation technique [10]. The degree of disorder or light-scattering properties of these systems were quantified in terms of the degree of light localization strength of these systems. The degree of light localization can be measured in terms of the mean inverse participation ratio <*IPR*> value of the optical eigenfunctions of the Anderson tight-binding model (TBM) Hamiltonian of the system, with closed boundary conditions. It is known that <*IPR*> value is proportional to the light localization properties of the system, and is a measure of the effective scattering strength of the system [11,12]. Therefore the <IPR($D_f$)> versus fractal dimension ($D_f$) are generated using the light localization technique to investigate the peak scattering point(s) or optimal scattering fractal dimension(s). Furthermore, we also studied random cut (*RC*) lattices, a non-fractal or pseudo fractal optical disordered samples/lattice media for <IPR($D_f$)> versus $D_f$ variation, to characterize optimal scattering points in 2D and 3D media, and to compare the results with the fractal samples generated by *DLA* method.

Schrodinger equation and Maxwell's wave equations can be projected to each other in all dimensions, as both the equations can be projected to Helmholtz equation with certain conditions, therefore, the developed methods of electronic systems can be used for optical systems [13]. We treat generated fractal optical samples using *DLA/RC* as optical lattice systems, with its lattice points as the scattering centers, by appointing a refractive index value for each lattice point, with two state values: lower (0) and higher (1). Different fractal geometries result in different distributions of the scattering centers with long-range or power-law correlations, which consequently leads to different degrees of disorder or disorder strengths. Once we generate lattices with different fractal dimensions, we use the Anderson-disorder tight binding model (TBM) with nearest neighbor interactions, under closed boundary conditions, a well-studied method in electronic system [14,15]. The Hamiltonian, *H*, for the system can be written as [15]:

$$H = \sum_i \varepsilon_i |i\rangle\langle i| + t \sum_{\langle ij \rangle} \{|i\rangle\langle j| + |j\rangle\langle i|\}. \quad (1)$$

In Eq. (1), $\varepsilon_i = (dn_i/n_0)$ is defined as the optical potential of $i^{th}$ site with $dn_i$ is the fluctuation above the mean refractive index $n_0$ [16], $t$ is the inter-lattice site hopping strength restricted to only the nearest neighbors. $|i\rangle$ and $|j\rangle$ are the optical wave functions at the $i^{th}$ and $j^{th}$ lattice sites, respectively, and <ij> indicates the nearest neighbors. The eigenvalues and eigenfunctions *(or $\Psi$ values)* of the system are obtained by diagonalizing the Hamiltonian *H*. Further, assuming the sample length *L* and the lattice size *a*, we can define the mean *IPR* value (<*IPR*>) over a sample as:

$$< IPR >_N = \frac{1}{N} \sum_{i=1}^{N} \int_0^L \int_0^L \Psi_i^4(x,y) dx dy, \quad (2)$$

where $\Psi_i$ is the $i^{th}$ eigenfunction of the Hamiltonian of the sample size *L*, $N=(L/a)^2$, $dx=dy=a$. Ensemble averaged *IPR*, <IPR>$_{\psi, ensemble}$ = <IPR>, is the degree of localization and proportional to the degree of disorder, the higher value of <IPR> implies the higher value of the disorder in the system [13]. For the Gaussian white noise disorder in 2D,

<IPR> = <<IPR>$_N$>$_{ensemble}$ ~ $dn \times lc$. $\quad (3)$

The fractal dimensions of the system, follow the scaling power law $N_r \propto r^{-D_f}$, with $D_f$ being the fractal dimension, and was taken by the well-known Minkowski–Bouligand dimension, or box-counting method. In this method, if a dimensional space containing a specific fractal structure is partitioned with a number of $N_r$-cubes $[N_r(r)]$ with side length *r*, then the $N_r$-cube counting fractal dimension is defined as [2,17]:

$$D_f = \lim_{r \to 0} [-\ln N_r(r)/\ln(r)] \quad (4)$$

Counting the number of $N_r$-cube of different sizes or values, the fractal dimension using equation (4) can be estimated by the slope of the linear fit of $ln(1/r)$ versus $ln(N(r))$ points. This method has proven accuracy of the fractal dimension calculation within ±0.05 for different deterministic fractals with known analytical fractal dimensions, such as Sierpinski's carpet and Sierpinski's triangle.

***Generation of fractal samples by DLA/RC simulations:*** In the following, we describe optical lattice generation by both the *DLA* (fractal) and *RC* (random cut/non-fractal) simulation methods.

***(i) Diffusion limited aggregation method (DLA):*** Fractal systems are generated by numerical simulations using a diffusion-limited aggregation (*DLA*) algorithm, a well-known process found in a wide array of fields such as biology, chemistry, material sciences, etc. In this technique, particles undergoing a random walk cluster together to form aggregates [10,18,19]. The rules for *DLA* fractal growths, in brief, are as follows: the simulation first creates a stationary 'seed' particle where the aggregation is then built by adding a second particle at some radius *r* away from the seed particle. The second particle undergoes a random walk until it encounters the seed particle, at which point it sticks with a certain probability, varying from 1 (full stick) to 0 (no stick) [10]. Lower sticky probability values allow the particles to 'fall' into gaps inside the lattice. If particle wonders a certain radius (kill radius) from the seed, the particle is destroyed and the process restarts, typical *DLA* generated samples are shown in Fig. 1. (a) and (b).

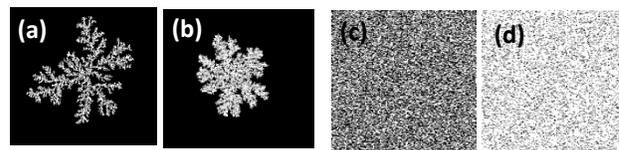

**Fig.1** Fractal structures generated by diffusion limited aggregation (*DLA*) simulation method in 2D with increasing stickiness parameter from left (a) to right (b). Numerically generated disorder matrix by random cut (*RC*) method simulation in 2D with randomly cutting lattIce points or randomly decreasing lattice points from left (c) to right (d).

***(ii) Random cut (RC) model:*** We also perform a simple random cut of lattice method starting from the net filled sample of filled lattice/matrix for disorder system, that does not have exact fractal correlation, for comparison with *DLA* method. In the *RC* method, 2D/3D filled matrices are generated by taking a lattice filled with points/particles, or scattering centers at every lattice point, and removing, or cutting the lattice points randomly. Two different stages of random cut are shown in Fig. 1(c)-(d), with lesser number of cuts (empty points) and a greater number of cuts from a filled lattice/matrix, respectively. This method does not generate a true fractal but rather a random disorder without long-range spatial correlation. The volume occupied ratio which is a simple measurement of how "full" the system is, rather than the calculated fractal dimension.

Once we generated the 2D/3D matrices, TBM Hamiltonians are formed for these matrices. We kept the onsite potential $\varepsilon$ constant at occupied position and 0 at non-occupied position throughout the lattice and the hopping parameter *t* as constant equal to 1. Simulations are performed in both 2D and 3D systems, using both *DLA* and *RC* techniques. 2D square lattices of length *L* = 8, 16, 32, and 64 corresponding to TBM Hamiltonian matrix size ($L^2$), N = 64, 256, 1024, and 4096 respectively are simulated. For 3D, cubic lattices of length *L*=4, 8, and 16 corresponding to TBM Hamiltonian matrix size of ($L^3$), *N* = 64, 512, and 4096, respectively, are constructed and simulated. In each complete iteration of a *DLA/RC* method, generated optical lattices are analyzed for its fractal dimension $D_f$ and average <IPR($D_f$)> value. In this method, we systematically scan the fractal dimension $D_f$ by box counting method and corresponding structural disorder strength by calculating $<IPR(D_f)>$. For *DLA* simulation, the arbitrary number of particles are simulated until the aggregation is built, and each aggregation is analyzed for its $D_f$ value and corresponding $<IPR(D_f)>$ value using Eqs. (1)-(4). For an expected system size $L$, the particles were added at a radius $r$ of twice $L$ with a kill radius of twice $r$, i.e. $2r$.

In the Hamiltonian, keeping the relative interaction strength ratio $\varepsilon/t$ in Eq. (1) as constant, the $<IPR(D_f)>$ vs $D_f$ curves are plotted for each system. The fractal dimension is adjusted directly for the *RC* method by controlling the number of particles cutting/removing from the lattice. The fractal dimension adjusted for the fractal system generated using *DLA* method is by controlling the stickiness parameter value. Since the largest stickiness value of 1 creates *DLA* fractals with larger values of the fractal dimension. The *DLA* and *RC* methods develop both 2D and 3D fractal systems that are simulated for eigenfunctions and then *IPR* value of a system. As can be seen in Fig. 2(a)-(b), $<IPR(D_f)>$ vs $D_f$ curves for different $\varepsilon/t$ constant values are shown for: (a) DLA generated fractal systems (b) RC generated fractal systems. These demonstrate a recognizable maximum, suggesting that there is indeed a fractal dimension to which optimal scattering occurs. This particular fractal dimension was dubbed the turning point, or $D_{ft}$. The turning point occurs for the RC method roughly when a maximum contrast between empty and occupied sites happens for RC systems. That is, at least, for RC systems, the turning points should occur when roughly half of the lattice sites are filled and half are empty. We found that for both the methods, the turning point showed dependency on systems length size $L$, as well as on the ratio $\varepsilon/t$ value. Increasing $L$ both shifts the turning point $D_{ft}$ towards a larger fractal dimension and stretches the plot in the vertical direction or higher localization. The averaged turning fractal dimension point for the 2D *RC* method with an energy ratio of 1 at a system length $L$ of 8, 16, 32, and 64 are 1.69, 1.79, 1.85 and 1.88 ±0.05, respectively. The vertical stretching of Fig.2(a)-(b) in $<IPR>$ is due to extremely high disorder-induced localization. The change of the energy ratio proved the most interesting in relation to the turning point. For 2D cases, both methods increasing $\varepsilon/t$ ratio correspond, expectantly, to an increase in the maximum $<IPR(D_f)>$ value as shown in Fig. 2(a) for *DLA* case and in Fig. 2(b) for the *RC* case. The maximum localization strength relative to the minimum can be described by the increase in onsite potential relative to the hopping parameter. As such, the maximum $<IPR(D_f)>$ values at low $\varepsilon/t$ energy ratio are not relatively strong compared to their minimal, and show instability below a ratio of 1, as the kinetic energy overwhelms the potential.

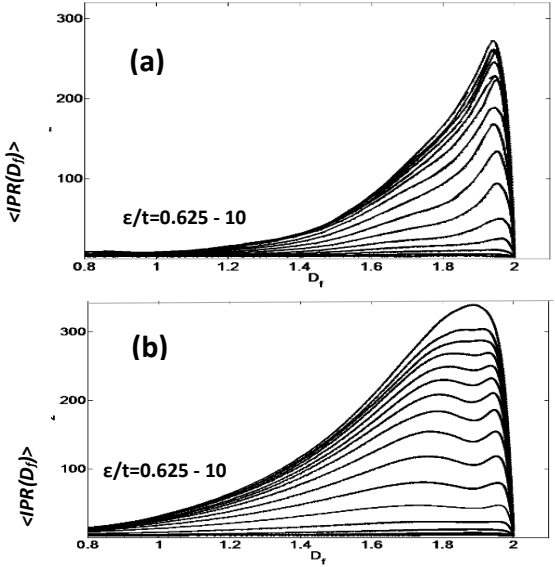

**Fig. 2**. 2D and lower fractal dimension $<IPR(D_f)>$ plots. Averaged $<IPR(D_f)>$ against $D_f$ plots for **(a)** *DLA* and **(b)** *RC* systems simulations shown with $L=64$, for different $\varepsilon/t$ values. Energy ratio $\varepsilon/t$ varied between 0.625 and 10 with intervals of 0.625; large ratios correspond to larger maximum $<IPR>$ values.

In Fig. 3(a)-(b), for 3D $<IPR(D_f)>$ vs $D_f$ curves are plotted for *DLA* and *RC*. The 3D systems present somewhat similar findings to 2D as shown for *DLA* [Fig. 3(a)] and *RC* [Fig. 3(b)]. The value of *IPR* is lower in 3D, suggesting weaker localization in 3D. There are two turning points, however, remains indefinitely regardless of how large the energy ratio becomes, for both DLA and RC fractals. Fig. 3(a), for DLA systems, shows curves are similar to its 2D counterpart, a local maximum $<IPR>$ value manifests at fractal dimension $D_{ft}= 2.24$ for all $\varepsilon/t >3$ and the second turning point $D_{ft}$ at 2.95±0.05. In Fig. 3(b), curved for RC method is presented. The saturating dimensions for two turning points are $D_{ft} = 2.24$ and 2.95±0.05, and is again symmetrical to the number of particles present in the system. The relative *IPR* strength of the double turning points, i.e. the optimal points/peaks, is much larger than the valley compared to its 2D counterpart as seen in Fig. 2(b). $\varepsilon/t$ value below 1 display extreme instability turning points for 3D systems compared to their 2D counterparts, as the system is less localized.

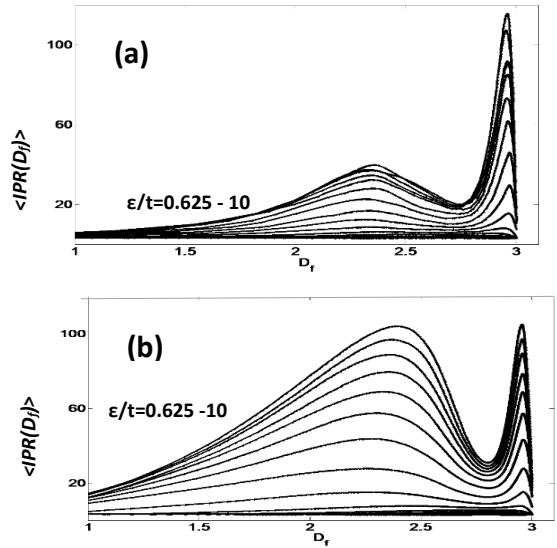

**Fig. 3.** 3D and lower fractal dimension $<IPR(D_f)>$ plots. Averaged $<IPR(D_f)>$ against $D_f$ plots for **(a)** *DLA* and **(b)** *RC* systems simulation shown for $L=16$ and different $\varepsilon/t$ values. Energy ratio $\varepsilon/t$ varies between 0.625 and 10 with intervals of 0.625; large ratios correspond to larger maximum $<IPR>$

In figure 4(a)-(b), we plot the $D_{ft}(\varepsilon/t)$ vs $\varepsilon/t$ curves for: (a) *DLA* and *RC* in 2D and (b) *DLA* and *RC* in 3D for constant system sizes. As shown in Fig. 4 (a), for 2D *RC* lattice, a bifurcation occurs for $D_{ft}(\varepsilon/t)$ vs $\varepsilon/t$ plot (□) for $\varepsilon/t$ value between 2 and 8, resulting in double turning points; the split in turning points is rather abrupt and temporary as one increases the energy ratio, before returning to its original turning point of $D_{ft}$=1.88±0.05, which corresponds to approximately half of the lattice sites filled. The fractals generated by *DLA* method (X) only ever singular turning point rises and then saturates after a ratio of around 3, to $D_{ft}$ =*1.94±0.05* that, unlike its *RC* counterpart, persists to a larger value of the ratio. This distinction between the two turning points of the *RC* method versus the singular for the *DLA* arises from the fact that, in terms of the *<IPR($D_{ft}$)>*, a random distribution will behave the same as the inverse of said distribution and as such, the turning points position of the *RC* method is symmetrical around the number of particles present and absent. For example, double turning points for 80% filling are corresponding to 20% and 80% of the lattice filled and these two peaks merge at 50%.

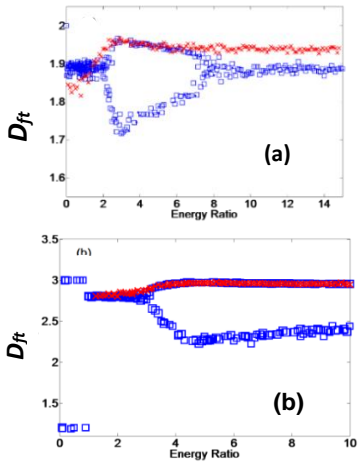

**Fig. 4.** Turning points $D_{ft}$ against energy ratio ε/t in 2D and 3D. **(a)** 2D system with size $L^2=(64)^2$, and **(b)** 3D system with size $L^3=16^3$. Blue squares represent *RC* method while red Xs (X) represent *DLA*.

Fig. 4(b) shows the turning point vs energy ratio plot, that is $D_{ft}(\varepsilon/t)$ vs $\varepsilon/t$ of energy ratio against the turning point for the 3D systems. For *RC* method, the double turning point values are shown (□), the saturation values are 2.24 and 2.95±0.05. The double turning points for the *DLA* is 2.24 (~constant value) and 2.95±0.05, are shown by symbol (X).

In conclusion, the inverse participation ratio (*IPR*) coupled with fractal analysis provides remarkable analysis technique for fractal disordered systems. The method provides the scattering properties of the system in a single parameter *<IPR>* value. <IPR($D_f$)> vs $D_F$ curve is non-monotonous for 2D/3D. The maximum/optimal scattering strengths occur at certain fractal dimension(s). This is because of the large separation of energy states that are symmetrical for random constant potentials of complementary particle lattice percentage occupation (for 80% filling, it is 80% and its complementary 20%). We have extended our results from weak to strong disorder to probe the optimal localization points.

**Importance of a turning point or optimal scattering fractal dimension ($D_{ft}$) in biological systems for cancer detection:**

Most biological systems are fractal, weakly disordered optical media, therefore, for biological systems ε/t ≤ 1 values are more relevant for light scattering from these systems. It can be seen from the simulation results that for a fractal disordered optical media, the localization properties of the systems are quite similar related to the turning points, or optimal points. That is, approximately the same $D_{ft}$ for different ε/t values. It is known that tissues/cells are fractal disordered optical media. The light scattering properties of tissues are important to understand the abnormalities in these tissues, such as progress of cancer. The structural abnormalities or alterations progress with the progress of diseases such as carcinogenesis or brain abnormalities [5,20]. There is no clear study to indicate which are the optimal scattering fractal dimensions in 2D or 3D tissues. This means, due to the non-monotonous nature of the fractal dimension in scattering, the degree of scattering can increase or decrease with the change of the fractal dimension, depending on the position of the fractal turning point $D_{ft}$ values. Therefore, it depends on the initial tissue structure or its fractal dimension and the direction of the change of the fractal dimension.

**Fundings.** Partial support by National Institutes of Health (NIH) grants (R01EB003682 and R01EB016983) and Mississippi State University to Dr. Pradhan are acknowledged.

**Disclosures**. The authors declare no conflicts of interest.

[1]  P. W. Anderson, Phys. Rev. **109**, 1492 (1958).
[2]  B. B. Mandelbrot, A. Blumen, M. Fleischmann, D. J. Tildesley, and R. C. Ball, Proceedings of the Royal Society of London. A. Mathematical and Physical Sciences **423**, 3 (1989).
[3]  C. M. Sorensen, Aerosol Science and Technology **35**, 648 (2001).
[4]  J. W. Baish and R. K. Jain, Cancer Res. **60**, 3683 (2000).
[5]  S. S. Cross, The Journal of Pathology **182**, 1 (1997).
[6]  D. S. Coffey, Nat Med **4**, 882 (1998).
[7]  S. Bhandari, S. Choudannavar, E. R. Avery, P. Sahay, and P. Pradhan, Biomed. Phys. Eng. Express **4**, 065020 (2018).
[8]  H. Namazi and M. Kiminezhadmalaie, Computational and Mathematical Methods in Medicine (2015).
[9]  C. J. R. Sheppard, Opt. Lett., OL **32**, 142 (2007).
[10] T. A. Witten and L. M. Sander, Physical Review B **27**, 5686 (1983).
[11]  P. Sahay, H. M. Almabadi, H. M. Ghimire, O. Skalli, and P. Pradhan, Opt. Express, OE **25**, 15428 (2017).
[12] P. Sahay, A. Ganju, H. M. Almabadi, H. M. Ghimire, M. M. Yallapu, O. Skalli, M. Jaggi, S. C. Chauhan, and P. Pradhan, Journal of Biophotonics **11**, e201700257 (2018).
[13] P. Pradhan and S. Sridhar, Phys. Rev. Lett. **85**, 2360 (2000).

[14] V. N. Prigodin and B. L. Altshuler, Phys. Rev. Lett. **80**, 1944 (1998).
[15] P. A. Lee and T. V. Ramakrishnan, Rev. Mod. Phys. **57**, 287 (1985).
[16] P. Pradhan and N. Kumar, Phys. Rev. B **50**, 9644 (1994).
[17] B. Dubuc, J. F. Quiniou, C. Roques-Carmes, C. Tricot, and S. W. Zucker, Phys. Rev. A **39**, 1500 (1989).
[18] E. Ben-Jacob and P. Garik, Nature **343**, 523 (1990).
[19] J. P. Gollub and J. S. Langer, Rev. Mod. Phys. **71**, S396 (1999).
[20] D. S. Coffey, Nat Med **4**, 882 (1998).

1. P. W. Anderson, "Absence of Diffusion in Certain Random Lattices," Phys. Rev. 109, 1492–1505 (1958).
2. B. B. Mandelbrot, A. Blumen, M. Fleischmann, D. J. Tildesley, and R. C. Ball, "Fractal geometry: what is it, and what does it do?," Proceedings of the Royal Society of London. A. Mathematical and Physical Sciences 423, 3–16 (1989).
3. C. M. Sorensen, "Light Scattering by Fractal Aggregates: A Review," Aerosol Science and Technology 35, 648–687 (2001).
4. J. W. Baish and R. K. Jain, "Fractals and cancer," Cancer Res. 60, 3683–3688 (2000).
5. S. S. Cross, "Fractals in Pathology," The Journal of Pathology 182, 1–8 (1997).
6. D. S. Coffey, "Self-organization, complexity and chaos: The new biology for medicine," Nat Med 4, 882–885 (1998).
7. S. Bhandari, S. Choudannavar, E. R. Avery, P. Sahay, and P. Pradhan, "Detection of colon cancer stages via fractal dimension analysis of optical transmission imaging of tissue microarrays (TMA)," Biomed. Phys. Eng. Express 4, 065020 (2018).
8. H. Namazi and M. Kiminezhadmalaie, "Diagnosis of Lung Cancer by Fractal Analysis of Damaged DNA," https://www.hindawi.com/journals/cmmm/2015/242695/.
9. C. J. R. Sheppard, "Fractal model of light scattering in biological tissue and cells," Opt. Lett., OL 32, 142–144 (2007).
10. T. A. Witten and L. M. Sander, "Diffusion-limited aggregation," Physical Review B 27, 5686–5697 (1983).
11. P. Sahay, H. M. Almabadi, H. M. Ghimire, O. Skalli, and P. Pradhan, "Light localization properties of weakly disordered optical media using confocal microscopy: application to cancer detection," Opt. Express, OE 25, 15428–15440 (2017).
12. P. Sahay, A. Ganju, H. M. Almabadi, H. M. Ghimire, M. M. Yallapu, O. Skalli, M. Jaggi, S. C. Chauhan, and P. Pradhan, "Quantification of photonic localization properties of targeted nuclear mass density variations: Application in cancer-stage detection," Journal of Biophotonics 11, e201700257 (2018).
13. P. Pradhan and S. Sridhar, "Correlations due to Localization in Quantum Eigenfunctions of Disordered Microwave Cavities," Phys. Rev. Lett. 85, 2360–2363 (2000).
14. V. N. Prigodin and B. L. Altshuler, "Long-Range Spatial Correlations of Eigenfunctions in Quantum Disordered Systems," Phys. Rev. Lett. 80, 1944–1947 (1998).
15. P. A. Lee and T. V. Ramakrishnan, "Disordered electronic systems," Rev. Mod. Phys. 57, 287–337 (1985).
16. P. Pradhan and N. Kumar, "Localization of light in coherently amplifying random media," Phys. Rev. B 50, 9644–9647 (1994).
17. B. Dubuc, J. F. Quiniou, C. Roques-Carmes, C. Tricot, and S. W. Zucker, "Evaluating the fractal dimension of profiles," Phys. Rev. A 39, 1500–1512 (1989).
18. E. Ben-Jacob and P. Garik, "The formation of patterns in non-equilibrium growth," Nature 343, 523–530 (1990).
19. J. P. Gollub and J. S. Langer, "Pattern formation in nonequilibrium physics," Rev. Mod. Phys. 71, S396–S403 (1999).
20. D. S. Coffey, "Self-organization, complexity and chaos: The new biology for medicine," Nat Med 4, 882–885 (1998).